\documentclass[twocolumn,showpacs,preprintnumbers,amsmath,amssymb]{revtex4}

\usepackage{graphicx}
\usepackage{dcolumn}
\usepackage{bm}
\usepackage{multirow}
\usepackage{amsmath}
\usepackage{hyperref}

\newcommand{\znsr}{$_{\mbox{\scriptsize Zn}}$}
\newcommand{\osr}{$_{\mbox{\scriptsize O}}$}

\newcommand{\astzn}{As$_{\mbox{\scriptsize Zn}}$-2V$_{\mbox{\scriptsize Zn}}$}
\newcommand{\aszn}{As$_{\mbox{\scriptsize Zn}}$-V$_{\mbox{\scriptsize Zn}}$}
\newcommand{\muzn}{$\mu_{\mbox{\scriptsize Zn}}$}
\newcommand{\muzno}{$\mu_{\mbox{\scriptsize ZnO}}$}
\newcommand{\muas}{$\mu_{\mbox{\scriptsize As}}$}
\newcommand{\muaso}{$\mu_{\mbox{\scriptsize As}_2\mbox{\scriptsize O}_3}$}
\newcommand{\muo}{$\mu_{\mbox{\scriptsize O}}$}

\begin{document}
\title{Defect energetics and electronic structures of As-doped {\it p}-type ZnO crystals: A first-principles study}

\author{Chol-Jun Yu, Yong-Guk Choe, Son-Guk Ri, and Myong-Il Kim}
\affiliation{Faculty of Materials Science, Kim Il Sung University, Ryongnam-Dong, Taesong District, Pyongyang, DPR Korea}
\author{Song-Jin Im}
\affiliation{Faculty of Physics, Kim Il Sung University, Ryongnam-Dong, Taesong District, Pyongyang, DPR Korea \\}

\date{\today}

\begin{abstract}
First-principles calculations based on density functional theory have been carried out to understand the mechanism of fabricating As-doped {\it p}-type ZnO semiconductors. It has been confirmed that \astzn~complex is the most plausible acceptor among several candidates for {\it p}-type doping by computing the formation and ionization energies. The electronic band structures and atomic-projected density of states of \astzn~defect complex-contained ZnO bulks have been computed. The acceptor level in \astzn~band structure has found to be 0.12 eV, which is in good agreement with the experimental ionization energy (0.12$\sim$0.18 eV). The hybridization among O 2p, Zn 3d and As 4s states has been observed around the valence band maximum.
\end{abstract}


\maketitle

\section{\label{intro}Introduction}
In recent years, ZnO-based materials have attracted intense attention due to their potential applications, for instance in ultraviolet light-emitting diodes, spintronics and lasing devices, as has been reviewed recently~\cite{Oezguer,Liua}. The attention is mainly motivated by attractive properties of ZnO bulk crystal with wurtzite structure such as direct wide band gap ($\sim$3.3 eV at 300 K) and large exciton binding energy ($\sim$60 meV). Moreover the crystal growth technology of ZnO is much simpler than for other wide band gap semiconductors like GaN ($\sim$3.4 eV at 300 K), resulting in a lower cost for ZnO-based devices.

The first-principles density functional theory (DFT) calculations of electronic band structure within the generalized gradient approximation (GGA) and local density approximation (LDA) with the mean-field Hubbard potential (LDA+{\it U}) approach~\cite{King,Zhang,Karazhanov1,Karazhanov2,Fisker} have revealed various electronic properties of ZnO bulk crystal. LDA+{\it U} calculation as a function of {\it U} and {\it J} has been found to correct the energy location of the Zn-3{\it d} electron levels and associated band parameters as well as the band gap in good agreement with the experimental data, while as common in semiconductors LDA underestimates the band gap and splitting energy between the states at the top of valence band(VB), but overestimates the crystal-field splitting energy~\cite{Karazhanov1,Karazhanov2}. The evolution of the bonding mechanism of ZnO under isotropic compression was studied by using full potential linearized augmented plane wave (FP-LAPW) method in LDA+{\it U}~\cite{Zhou} and the honeycomb structures was investigated~\cite{Topsakal}. The optical absorption and excitonic properties of wurtzite ZnO were investigated by means of the first-principles approach taking into account electron-hole correlations, focusing on the calculation of the band edge optical spectra. The calculated exciton binding energies are around 68 meV~\cite{Laskowski,Zhukov}. 

As in any semiconductor, defects affect the electrical and optical properties of ZnO. Since Dietl {\it et al}~\cite{Dietl} predicted that GaN- and ZnO-based diluted magnetic semiconductors (DMS) could exhibit ferromagnetism above room temperature when doped with transition metal (TM) elements in {\it p}-type materials, it has been reported experimentally that bulk Mn and Co-doped ZnO crystals show room temperature ferromagnetism for TM concentrations of 1 at. \%~\cite{Polyakov,Trolio,Ivill,Yan,Sharma,Heyd,Sanyal}. Through the first-principles DFT electronic structure calculations of TM-doped ZnO DMS~\cite{Petit,Singh,Sandratskii,Raebiger}, specially Mn-doped~\cite{Shi,Sharma,Feng,Wang}, V-doped~\cite{Yu}, and Co-doped~\cite{Liu,Patterson,Lany,Iusan,Chanier,Hu,Trolio,Sanyal} materials, it was found that the strong {\it s-d} as well as {\it p-d} hybridization play a dominant role in such ferromagnetism, owing to small nearest neighbor distance and small spin de-phasing {\it spin-orbit} interaction. Local spin density approximation(LSDA)+{\it U} that takes into account the strong electron correlation gives the reasonable results in agreement with the experimental one, while LSDA predictions might be misleading and should be considered with care~\cite{Iusan,Chanier,Hu}.

Another way in searching the functional doped ZnO is to make easy the fabrication of {\it p}-type ZnO by doping non-transition metal~\cite{Limpijumnong,Zhang} in the place of Zn and N in the place of O. From the first-principles calculations it was shown that the oxygen vacancy V\osr~is not a shallow donor, but has a deep $\varepsilon$(2+/0) level at 1.0 eV below the conduction band\cite{Janotti,Paudel}. Meanwhile it has been revealed that the level of the nitrogen atom on the oxygen site, N\osr, is relatively deep, making acceptor ionization difficult, and N-doped ZnO could be unstable~\cite{Park}. By the first-principles plane wave ultrasoft(US) pseudopotential calculations, Limpijumnong {\it et al}~\cite{Limpijumnong} have calculated the formation and ionization energies of possible As and Sb defect complexes, and concluded that among several defect complexes \astzn~is the most plausible because of its low formation energy of $\sim$1.59 eV and low ionization energy of $\sim$0.15 eV, reasonable value compared with the experimental value ($\sim$0.12 eV). Du {\it et al}~\cite{Du} confirmed this prediction by fabricating the ZnO/GaAs heterojunction using metal-organic chemical vapor deposition(MO-CVD) and by measuring the visual-infrared electroluminescence emission.

In this paper we have calculated the defect formation energetics and presented the electronic structures of As-doped defect complexes in order to refine the mechanism of fabricating {\it p}-type ZnO semiconductors. Firstly the defect energetics of As defect complexes have been calculated using ultrasoft(US) pseudopotential plane wave method in LDA, and then the electronic band structures with DOSs have been investigated using norm-conserving(NC) pseudopotential pseudo atomic orbital(PAO) method in spin-polarized GGA.

\section{Computational Method}
\begin{figure}[!ht]
\begin{center}
\includegraphics[clip=true,scale=0.4]{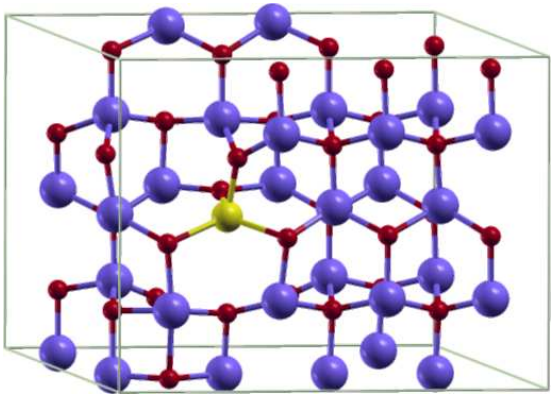}
\caption{\label{fig:supercell}(Color online)64 atom supercell for \astzn~defect complex, where pink circle for zinc, red circle for oxygen, and yellow circle for arsenic atoms, respectively.}
\end{center}
\end{figure}

Our research work was based on state-of-the-art first-principles calculations using DFT, pseudopotential, and plane wave and pesudoatomic orbital (PAO) expansion of wave functions and potentials. Supercell geometries with 64 atoms were employed, as typically shown in figure \ref{fig:supercell}. Experimentally determined lattice parameters for wurtzite ZnO crystal (a=3.253 \AA, c=5.209 \AA, u=0.383) have been used in the present first-principles calculations.

The formation energy of a defect D in charge state $q$ is defined as~\cite{Limpijumnong}
\begin{equation}
\label{eq:defectene}
\begin{aligned}[b]
E^f(D^q)=&E_{tot}(D^q)-E_{tot}(bulk)+\Delta n_{\mbox{\scriptsize Zn}}\mbox{\muzn} \\
 &+\Delta n_{\mbox{\scriptsize O}} \mbox{\muo}+\Delta n_{\mbox{\scriptsize As}}\mbox{\muas} + qE_F,
\end{aligned}
\end{equation}
where $E_{tot}(D^q)$ and $E_{tot}(bulk)$ are the total energies of the supercell with and without the defect, $\Delta n_{\mbox{\scriptsize A}}$ and $\mu_{\mbox{\scriptsize A}}$ are the number of species A (=Zn, As, O) removed from a perfect crystal to its respective reservoir and the corresponding reservoir chemical potential, respectively, and $E_F$ is the Fermi energy with respect to the valence band maximum(VBM). The chemical potentials depend on the growth conditions. The maximum values of \muzn, \muas, and \muo~are the total energies of Zn metal, As solid, and O$_2$ gas, respectively. However, \muzn~and \muo~are not independent, but related as \muzn + \muo$\leqslant$\muzno. Such relation is also hold for \muas~and \muo; 2\muas+3\muo$\leqslant$\muaso.

A lower formation energy implies a high equilibrium concentration of the defect, while a high formation energy means that defects are not plausible to form. The concentration of the defect in a crystal depends upon its formation energy as follows,
\begin{equation}
\label{eq:concent}
 c=N_{sites} \exp\left(-\frac{E^f}{k_BT}\right),
\end{equation}
where $N_{sites}$ is the number of defect sites in the crystal.

For more reliable estimation of defect formation energetics, we have employed US pseudopotentials~\cite{vanderbilt} in LDA~\cite{pw91}, 300 eV as a cutoff energy for the plane wave expansion, and $(2\times4\times2)$ Monkhorst-Pack~\cite{Monk-pack} $k-$points mesh without shift for Brillouin zone integration. Atoms were fully relaxed until the forces converge to 0.01 eV/\AA. The calculations were done using Quantum-ESPRESSO package~\cite{pwscf}.

Then we have calculated the electronic band structures and DOSs in GGA (revised PBE form)~\cite{rpbe} and spin-polarized cases, which were carried out using pseduopotential PAO method implemented in SIESTA code~\cite{siesta}. In the construction of pseudopotential of Zn, 3d electrons are included as the valence electrons. The configurations of valence electrons were $4s^23d^{10}4p^0$ in Zn, $2s^22p^4$ in O, and $4s^24p^3$ in As, respectively. The PAO basis sets of the standard double zeta plus polarization (DZP) with energy shift of 10 meV were used. The mesh cutoff energy was 200 Ry, and the kgrid cutoff for generating {\it k}-points was 10 \AA.

\section{Results and discussion}
\subsection{Defect complexes}
As mentioned above, it is not easy to fabricate {\it p}-type doping in ZnO, because of (1) the compensation of dopants by low-energy native defects like Zn$_i$ or V\osr, (2) low solubility of the dopant like N\osr, and (3) deep impurity level causing significant resistance to form shallow acceptor level.

The possible acceptors in ZnO are known as one-valence electron elements, for instance alkali metal elements, Cu and Ag, and Zn vacancies (V\znsr) on the two-valence electron element Zn sites, and five-valence elements like N, P, As, and Sb on the six-valence element oxygen sites, because of the lack of one electron by such substitutions. Due to deep acceptor levels, however, many of these defects did not contribute significantly to {\it p}-type conduction. In terms of shallowness of acceptor levels alkali elements on Zn sites could be better {\it p}-type dopants than group-V elements on oxygen sites. However, alkali elements tend to occupy the interstitial sites rather than substitutional sites due to partly their small atomic radii and more electronegativity (see Table \ref{tab:ioninfo}), and thus act mainly as donors ({\it n}-type dopants) instead of acceptors. Moreover, remarkably larger bond length for Na and K than Zn--O bond length induces lattice strain, increasingly forming vacancies that compensate the dopants. Similar large size mismath with oxygen is seen for group-V elements, except for nitrogen, which also have significatly larger bond lengths and thus are more likely to form antisites to avoid the lattice strain. Therefore, the most promising candidate for {\it p}-type dopants may be nitrogen on the oxygen sites as it has similar atomic size and electronegativity, and it does not form the N\znsr~antisite. However, it is well known that the solubility of N in ZnO is low. Therefore, a donor-acceptor codoping method that can enhance the solubility of N has been proposed and both experimental~\cite{Du} and theoretical~\cite{Limpijumnong} studies have been performed.
\begin{table}
\caption{\label{tab:ioninfo}Ionic and covallent radii, nearest-neighbor bond (with oxygen) lengths in the unit of \AA~and electronegativities of the elements for dopants in ZnO. The values in brakets present the ionic charge values and Elec. stands for electronegativity. ($^a$ from Ref\cite{Tilley}, $^b$from Ref\cite{Park} and $^c$ from Ref\cite{Rohrer}.)}
\begin{ruledtabular}
\begin{tabular}{clccc}
Element & Ionic$^a$ & Covallent & Bond $^b$ & Elec.$^c$ \\
\hline
O  & 1.26 (-2)            & 0.66 &  --  & 3.44 \\
N  & 0.02 (+3)            & 0.70 & 1.88 & 3.04 \\
P  & 0.31 (+3), 0.56 (+5) & 1.09 & 2.18 & 2.19 \\
As & 0.64 (+3)            & 1.20 & 2.23 & 2.18 \\
Sb & 0.75 (+3)            & 1.37 &      & 2.05 \\
\hline
Zn & 0.89 (+2)            & 1.39 & 1.95 & 1.65 \\
Cu & 1.08 (+1), 0.87 (+2) & 1.28 &      & 1.90 \\
Ag & 1.29 (+1)            & 1.44 &      & 1.93 \\
Li & 0.88 (+1)            & 1.56 & 2.03 & 0.98 \\
Na & 1.16 (+1)            & 1.91 & 2.10 & 0.93 \\
K  & 1.52 (+1)            & 2.34 & 2.42 & 0.82 \\
\end{tabular}
\end{ruledtabular}
\end{table}

\begin{figure}[!ht]
\begin{center}
\includegraphics[clip=true,scale=0.43]{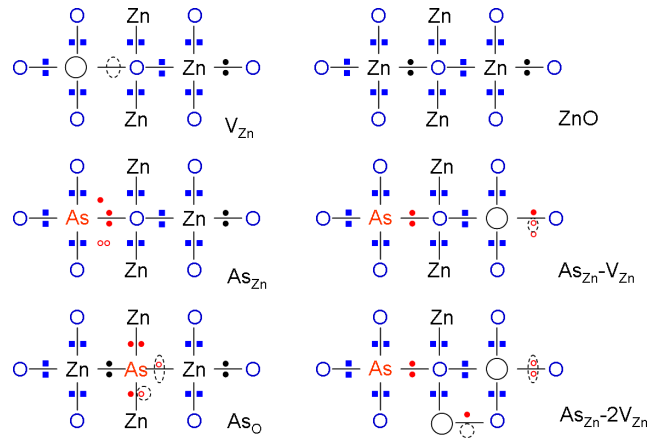}
\caption{\label{fig:defect-scheme}(Color online) Schematic figure showing how As-doped defect complexes could be acceptor or doner. Zn and O atoms provide two (dark circles) and six (blue squares) electrons respectively, and As atom can give three (red circles) and additionally two electrons (red open circles), to form covallent bonds.}
\end{center}
\end{figure}

When As doped into ZnO bulk, there will be several defects such as As\znsr~and As\osr, and also codoped defects like \aszn~and \astzn. Because As can give either three 4p electrons or five 4s4p electrons, there are extra either one or three electrons if As-O bond is formed in the place of Zn-O bond, and thus As\znsr~is surely a doner. Meanwhile, when six valent O is subsitituted by As, either one or three electrons are of lack and thus As\osr~is a acceptor. Moreover, As-codoped defect complex \astzn~can also be a acceptor because of the lack of one or three electrons, whereas \aszn~could be either a acceptor or a doner depending on how many electrons can provide, as depicted in figure \ref{fig:defect-scheme}.
We have calculted the formation energies of As-doped and codoped defect complexes in ZnO. By using US pseudopotentials the DFT total energies of wurtzite-type ZnO, orthohombic As$_2$O$_3$(a=7.043 \AA, b=5.154 \AA, c=5.393 \AA, 4-atomic units) units and O$_2$ gas molecule (using a=12 \AA~cubic supercell) were $-10.16$ eV, $-31.99$ eV and $-8.81$ eV, from which we can obtain the chemical potentials of the elements as \muzn=$-5.75$ eV, \muas=$-9.39$ eV and \muo=$-4.41$ eV, respectively.

Table \ref{tab:defectene} shows the calculated defect formation energies, which are overally agreed with the previous first-principles calculations. Note that we did not consider the oxygen-rich conditions, i.e., \muo= 0, but normal condition \muo=$-4.41$ eV, while using $E_F$=0. In such conditions, the formation energies of doner defects are in general low, even negative values showing spontaneous process in the case of isolated As defect, As\znsr (3+). From these result it is again said that the fabrication of {\it n}-type doping in ZnO is relatively easier than {\it p}-type doping.
\begin{table}
\caption{\label{tab:defectene}Calculated defect formation energies, where Acc. and Don. stand for acceptor and doner, respectively. The values in brackets are from Ref\cite{Limpijumnong}. }
\begin{ruledtabular}
\begin{tabular}{lcccc}
Defect  & Charge & State & \multicolumn{2}{c}{$E^f$ (eV)} \\
\cline{4-5}
        &        &       & This work & Ref\cite{Limpijumnong}\\
\hline
\astzn  & 0      &       & 1.83 & 1.59      \\
        & $-$    & Acc.  & 1.98 & 1.74      \\
        & $3-$   & Acc.  & 4.59 & 4.47      \\
\aszn   & $+$    & Don.  & 0.22 & 0.81      \\
        & 0      &       & 3.45 &           \\
        & $-$    & Acc.  & 2.95 & 3.06       \\
As\znsr & $3+$   & Don.  &$-2.50$ & $-0.68$      \\
        & $+$    & Don.  & 0.28 & 1.23      \\
        & 0      & Don.  & 2.75 &            \\
As\osr  & 0      & Acc.  & 10.21 & 9.87     \\
        & $-$    & Acc.  & 11.75 & 10.80    \\
V\znsr  & 0      & Acc.  & 1.99 & 1.86      \\
        & $2-$   & Acc.  & 2.63 & 2.63      \\
\end{tabular}
\end{ruledtabular}
\end{table}

Here we focus on the acceptor defects: Isolated As defect on the oxygen sites has the highest formation energy among them in one hand, and therefore it is again clear that the realization of such defect formation is very difficult. The reason of this high formation energy is mainly due to the significantly large size mismatch between oxygen and arsenic atoms. On the other hand, As-codoped defect complex, \astzn~has the lowest formation energy of 1.83 eV, which is slightly higher than the previous value of 1.59 eV. The binding energy of (As\znsr--V\znsr)$^+$ from As$_{\mbox{\scriptsize Zn}}^{3+}$~and V$_{\mbox{\scriptsize Zn}}^{2-}$~is $-0.09$ eV, thus saying endothermic rather than exothermic process as in the previous result~\cite{Limpijumnong}. This is due to much low formation energy of As$_{\mbox{\scriptsize Zn}}^{3+}$~defect calculated in the present work. The binding energy of (As\znsr--V\znsr)$^-$ from (As\znsr--2V\znsr)$^-$~and V$_{\mbox{\scriptsize Zn}}^{2-}$~is $0.87$ eV, still lower than the previous result 1.70 eV. The total binding energy of (As\znsr--2V\znsr)$^-$ from As$_{\mbox{\scriptsize Zn}}^{3+}$~and 2V$_{\mbox{\scriptsize Zn}}^{2-}$ is 0.78 eV. The ionization energy of \astzn, $\varepsilon (0/-)$, is 0.15 eV, which is exactly same to the previous first-principles calculation and is in good agreement with the experimental value (0.12$\sim$0.18 eV).

\subsection{Electronic structure}
\begin{figure}[!t]
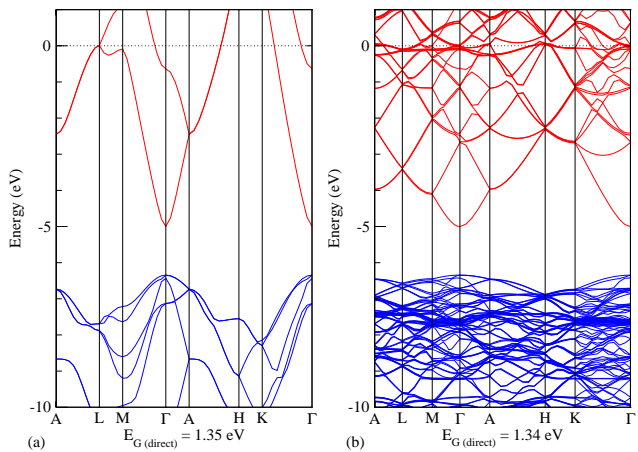

\begin{center}
\includegraphics[clip=true,scale=0.35]{fig3-a.eps}
\includegraphics[clip=true,scale=0.35]{fig3-b.eps}
\caption{\label{fig:band-ZnO}(Color online)Band structures of perfect ZnO bulk in the wurtzite primitive cell (a) and $4\times2\times2$ supercell (b).}
\end{center}
\end{figure}
We then calculated the electronic band structure of above-mentioned As-doped ZnO bulk materials using pseudopotential PAO method implemented in SIESTA. We first have tested the PAO basis sets and xc functionals with perfect ZnO bulk. With other parameters like mesh cutoff energy and kgrid cutoff fixed as 200 Ry and 10 \AA--their changes leave the bandgap almost invariant--the DZP in GGA (RPBE) provided the biggest bandgap as 1.35 eV, which is considerably underestimated value in the comparisons with both experimental (3.47 eV) and explicite many-body (GW or LDA+{\it U}) computational values ($\sim$3.34 eV). Because in this work we intend to investigate the general tendency in the electronic structure of above-mentioned As-doped ZnO bulk rather than the exact estimation of bandgap, we have left this underestimation of bandgap on the shelf. The band structure of ZnO bulk in wurtzite primitive unit cell using the DZP PAO basis sets and RPBE GGA xc functional with spin-polarization is shown in figure \ref{fig:band-ZnO}.

We have calculated the band structures and the density of states (DOS), atomic-projected partial DOS (PDOS), of the supercells with and without defects. For the calculation of DOS we set the histogram parameter as 0.2 eV and the energy interval as $[-26, 15]$eV with the total number of energy points of 1000. The atomic coordinates optimized by the previous US plane wave calculations for the defect energetics were adopted.
\begin{figure}[!t]
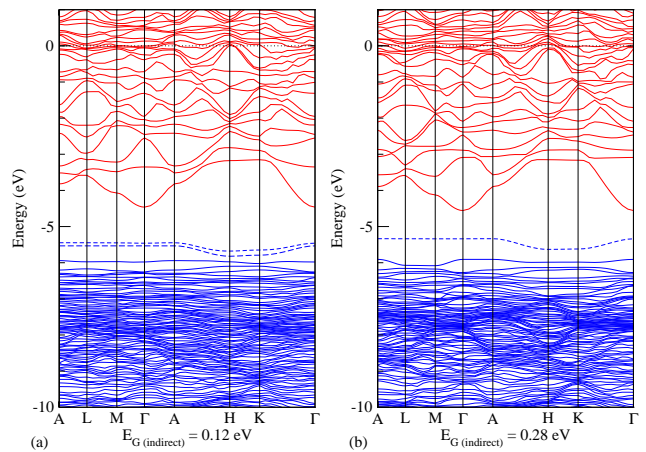

\begin{center}
\includegraphics[clip=true,scale=0.35]{fig4-a.eps}
\includegraphics[clip=true,scale=0.35]{fig4-b.eps}
\caption{\label{fig:band-aszn-vzn}(Color online)Band structures of (As\znsr--2V\znsr)$^{3-}$~(a) and (As\znsr--V\znsr)$^-$~(b). The dashed lines indicate the additional level, which can be thought as acceptor level.}
\end{center}
\end{figure}

To show the acceptor level, we plot the band structures of (As\znsr--2V\znsr)$^{3-}$~and (As\znsr--V\znsr)$^-$ defect complexes-contained bulks in figure \ref{fig:band-aszn-vzn}. We could see that on top of the VBM there is a level of 0.12 eV for the former and a level of 0.28 eV for the latter, which can act as a acceptor level. Let us consider this subtle assertion in more detail. In our calculation we used pseudopotentials for all elements, whose number of valence electrons are 12, 6 and 5 for Zn, O and As, respectively. Therefore, the total numbers of valence electrons are 576 in the $4\times2\times2$ perfect supercell, 545 in \astzn~supercell, and 557 in \aszn~supercell. When spin degeneracy is considered, thus, the numbers of valence bands in these supercells are 288, 272.5 and 278.5. To simulate the charged supercells (of course, this can be done using a homogeneous charge background in the algorithm) we inserted additional electrons of one or three. To plot the band structures of these charged supercells, we regarded the number of valence bands as 272 for (As\znsr--2V\znsr)$^{3-}$~and 278 for (As\znsr--V\znsr)$^-$~supercells. In the former case one band is half-filled and, after insertion of three electrons, the band is filled and one additional band is appeared. On the other hand, a half-filled band is filled by insertion of one electron and thus no additional band is appeared in the latter case. This thought is reasonable compared with the real process of electron excitation in semiconductors.

Although the bandgap is in general underestimated within LDA or GGA, we allege boldly that the obtained bandgaps are reliable values; we have calculated the band energies of the charged supercells (with additional electrons) and, thus, thease energy gaps are not really bandgaps but a kind of energy gap between valence-like bands. The energy gap of 0.12 eV is close to the calculated ionization energy of 0.15 eV and experimental ionization energies of 0.12 and 0.18 eV. Moreover, the higher acceptor level of \aszn, 0.28 eV, indicates that \astzn~defect complex is better acceptor than \aszn~defect complex.

\begin{figure}[!ht]
\begin{center}
\includegraphics[clip=true,scale=0.6]{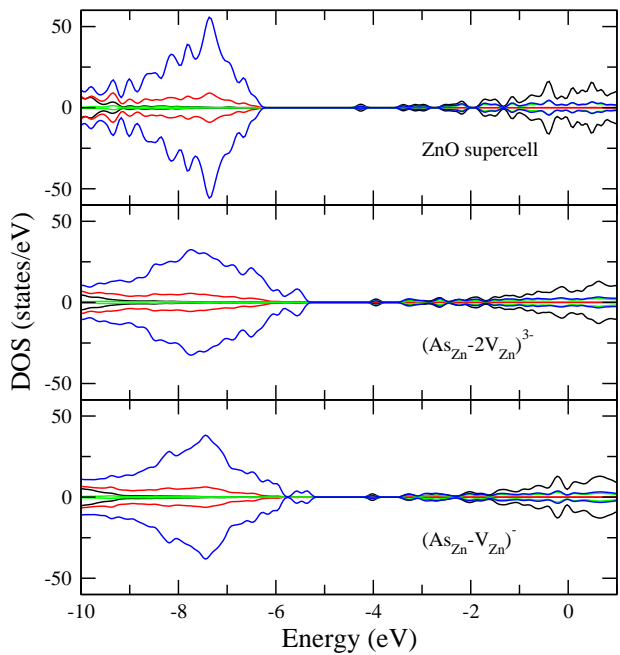}
\caption{\label{fig:pdos}(Color online)Atomic-projected density of states, in the unit of States/eV, for the perfect $4\times2\times2$ ZnO supercell, \astzn~and \aszn~defect complex-contained ZnO supercells. The black line is for Zn 4s, red line for Zn 3d, green line for O 2s and blue line for O 2p electrons, respectively.}
\end{center}
\end{figure}
Figure \ref{fig:pdos} shows the corresponding atomic-project DOSs. It is observed from the figure that no change between majority- and minority-spin PDOSs happens, thus saying that the magnetism is not found in these defect complexes in contrast to TM-doped ZnO materials. We can also find that around the VBM oxygen 2p electrons play a dominant role and next zinc 3d electrons play a certain role, while expecting strong hybridization between Zn 3d states and O 2p states. By comparing the bulk ZnO band structure with the As-doped ZnO bulk band structures in figure \ref{fig:pdos}, it can be seen that some of the oxygen 2p electrons can be excited to the acceptor level. We also find that such excitation of oxygen 2p electrons is more active in (As\znsr--2V\znsr)$^{3-}$~defect complex than in (As\znsr--V\znsr)$^-$~defect complex, because of the higher and wider peaks of PDOS around the VBM in the former complex.
\begin{figure}[!ht]
\begin{center}
\includegraphics[clip=true,scale=0.6]{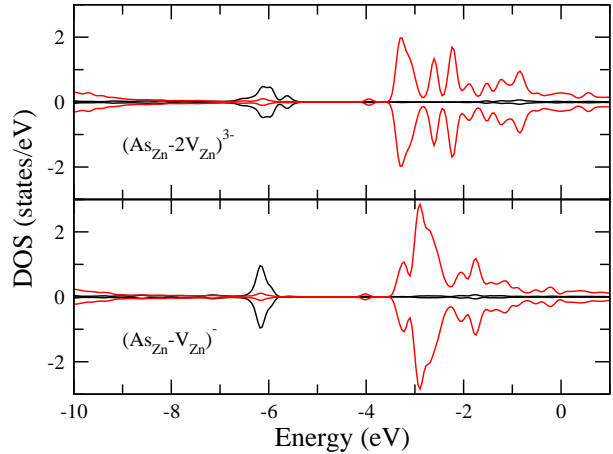}
\caption{\label{fig:asdos}(Color online)As atom-projected density of states, in the unit of States/eV, for \astzn~and \aszn~defect complex-contained ZnO supercells. The black line is for As 4s and red line for As 4p electrons, respectively.}
\end{center}
\end{figure}
Furthermore, figure \ref{fig:asdos} shows As atomic-projected DOS, where around the VBM As 4s electrons are dominant and As 4p electrons contribute to the conduction bands. Therefore, further hybridization from As 4s states to Zn 3d--O 2p hybridized state is expected. From the comparison between two PDOSs it can be seen that the hybridization from As 4s is stronger in \aszn~complex than in \astzn~complex. 

\section{Summary}
Using the DFT first-principles pseudopotential plane wave and PAO basis set method within LDA and GGA, the defect energetics and the electronic structures of As-doped wurtzite ZnO in $4\times2\times2$ supercell have been computed in order to get insight how to fabricate {\it p}-type ZnO semiconductors. The calculations of defect formation energies for possible As-doped ZnO models confirmed that \astzn~defect complex is the most prospect {\it p}-type doping with the lowest formation energy of 1.83 eV and moderate ionization energy of 0.15 eV. The band structures of (As\znsr--2V\znsr)$^{3-}$ and (As\znsr--V\znsr)$^-$ calculated pseudopotential PAO method in RPBE xc functional GGA contained the acceptor level at 0.12 eV, which is close to the calculated ionization energy and the experimentally measured values (0.12 $\sim$ 0.18 eV). From the analysis of atomic-projected density of states for these defect complexes, the hybridization among O 2p, Zn 3d and As 4s states play a certain role in the electron excitation.

\section*{\label{ack}Acknowledgments}
The work was partially supported from the Committee of National Science and Technology, DPR Korea under the grant number of 10-02-4 for national research project.

\bibliography{Reference}

\end{document}